# An Ontology-Based multi-domain model in Social Network Analysis: Experimental validation and case study

José Alberto Benítez-Andrades[a,*], Isaías García-Rodríguez[b], Carmen Benavides[a], Héctor Aláiz-Moretón[b], José Emilio Labra Gayo[c]

[a] SALBIS Research Group, Department of Electric, Systems and Automatics Engineering, University of León, Campus of Vegazana s/n, León, 24071, León, Spain {jbena@unileon.es, carmen.benavides@unileon.es}

[b] SECOMUCI Research Group, Escuela de Ingenierías Industrial e Informática, Universidad de León, Campus de Vegazana s/n, C.P. 24071 León, Spain {igarr@unileon.es, hector.moreton@unileon.es}

[c] Department of Computer Science, Universidad de Oviedo, C/Calvo Sotelo, s/n, 33007 Oviedo, Spain {labra@uniovi.es}

**Abstract:**

The use of social network theory and methods of analysis have been applied to different domains in recent years, including public health. The complete procedure for carrying out a social network analysis (SNA) is a time-consuming task that entails a series of steps in which the expert in social network analysis could make mistakes. This research presents a multi-domain knowledge model capable of automatically gathering data and carrying out different social network analyses in different domains, without errors and obtaining the same conclusions that an expert in SNA would obtain. The model is represented in an ontology called OntoSNAQA, which is made up of classes, properties and rules representing the domains of People, Questionnaires and Social Network Analysis. Besides the ontology itself, different rules are represented by SWRL and SPARQL queries. A Knowledge Based System was created using OntoSNAQA and applied to a real case study in order to show the advantages of the approach. Finally, the results of an SNA analysis obtained through the model were compared to those obtained from some of the most widely used SNA applications: UCINET, Pajek, Cytoscape and Gephi, to test and confirm the validity of the model.

Keywords Ontology-based systems; semantic web; semantic technologies; social network analysis; ontology multi-domain; knowledge-based systems.

**Highlights:**
- **The Social Network Analysis applications do not currently facilitate the different tests.**
- **It is possible to create an ontology within the scope of the multi-domain SNA.**
- **The proposed model has been validated by comparing the results with other validated tools.**
- **The Semantic Web applied to the SNA is a line of research with a high capacity for growth.**

# 1. Introduction:

Today, the term 'social network' is associated with a number of interactive tools that can be found on the Internet such as Facebook, Twitter or Instagram. However, especially in the area of social sciences, the term "social networks", makes reference to a finite set of actors and the relationships that link them [1]. These social networks are considered as social structures in which communication processes and transactions between people take place. Once the concept of a social network is defined, the Social Networks Analysis quantifies the relationships between the actors with the aim of creating matrix and graphic networks that represent these relationships as a whole, and thus analyzes the different characteristics of the system that is being studied, indistinctly of the type of relationship to which they belong: political, economic, kinship, friendship, cooperation, conflict, etc [2,3].

Social Network Analysis (SNA) techniques have mushroomed extensively in recent years in different areas of study within multiple domains. In the field of health, it applies both to socio-health management and studies into different diseases within the population [3]. Some examples applied to medicine include the study of affiliation networks of HIV health centers of men of color who have sexual relations with people of the same sex, enabling different patterns to be established and thus helping to prevent the disease in this population [4]. Another example in which the prevention of different sexually transmitted infections is also sought is that of Niekamp et al. [5] or a qualitative study of secondary school teachers' perception of Social Network Analysis metrics in the context of alcohol consumption among adolescents [6]. SNA is not only applied to health, this technique is also being applied in different domains, for example, the use of SNA to assess and predict future knowledge flows of an insurance company [7] or a study responsible for measuring the impact of Google Docs on student collaboration [8].

The complete procedure, which involves carrying out a Social Network Analysis, begins with the development of a questionnaire by the expert in Social Network Analysis. The subsequent data collection after its completion by a population group, modeling the data, the application of an SNA application to obtain global and local measurements [9–11] and finally the interpretation of the aforementioned measures by the SNA expert. This complete procedure involves a high temporary cost on the part of the SNA expert, especially when it comes to modeling the data collected to create the different social networks that you wish to analyze, as well as there being a high probability of making mistakes at some step which means potentially repeating the social network analysis on numerous occasions.

The automation of the Social Network Analysis procedure to reduce time and eliminate the appearance of errors is possible using any computational solution. But Semantic Technologies offer added value within the entire procedure that other technologies do not. Thanks to the semantic web, it is possible not only to automate the process for obtaining and collecting data together with the SNA measures [12], but also to obtain conclusions from an applied SNA automatically. Furthermore, as a result of the proposed problem, it has been found necessary to apply technologies that help the expert to automate the entire SNA procedure applied to any domain. For this reason, in the wake of the problems arising from the complete procedure of an SNA, this research proposes a solution using semantic technologies based on the creation of a multi-domain ontology in the field of SNA.

To date, solutions to SNAs using existing semantic technologies have not come up with a complete solution to the problem. The conceptual models found have facilitated compliance with the basic rules of the semantic web, which suggest the reuse of existing ontologies when trying to solve a problem. However, none of the solutions found is a complete solution.

The remainder of this article is arranged as follows: Section 2 is a literary review of the semantic technologies applied to the analysis of social networks; Section 3 describes the design of the research and the procedure for the creation of a multi-domain conceptual model in the area of SNA; Section 4 details the results obtained after the validation procedure of the generated model; Section 5 sets out the final conclusions from the experiments carried out throughout the research.

## 2. Literature review:

*2.1. Difficulties in Social Network Analysis studies*

Many of the researchers in different fields are in need of epidemiological studies and more specifically cross-sectional studies. These types of studies are based mainly on obtaining information from different groups of people through the use of surveys and then, curating the data and carrying out a social network analysis [13–15].

To date, researchers have carried out the surveys using different tools such as Google Docs [8], SurveyMonkey [16] or Survio.com. These tools are very useful because they allow you to create surveys in a simple way and obtain the data in different formats. It is possible to obtain this data in CSV format or on an Excel spreadsheet. However, after obtaining the data, the subsequent analysis must be carried out completely by the researchers. Often, the questionnaires really show data that are useful after their prior treatment (data curation).

Some of the validated questionnaires in the field of health are, for example, FAS II[1], AUDIT[2] [17], KIDSCREEN 27[3] [18] or EDADES [4]. These questionnaires are usually used to obtain information on the socioeconomic level of the target group (FAS II), as well as finding out the risk of alcohol consumption in the population (AUDIT and AGES) or to know the levels of personal satisfaction existing in different areas (KIDSCREEN). However, these questionnaires are made up of questions that have a value, and after the sum total of these values are calculated, a numerical result is obtained that represents a real risk or an area in which the respondent is located.

Many researchers currently have to carry out the collection and curation an analysis of data manually, finding different problems such as the excessive time used to handle the data obtained in the surveys, the appearance of a multitude of relative errors due to a mistake in the sum of the data by the researcher, etc. Some of these errors have been mentioned in different studies. One example is the study of Morris et al. Which mentions that the large amount of lost data in the collected data set is a serious problem, [19]. This problem caused difficulties in the subsequent SNA. Other problems found on this kind of study are, for example: (i) difficulty in being able to create new networks taking into account the values in the weighted relationships, (ii) errors in curing data produced by manually transferring the data, and (iv) difficulty in interpreting the results by not obtaining possible conclusions through prior analysis, among others [20].

---

[1] FAS II: Family Affluence Scale II
[2] AUDIT: Alcohol Use Disorders Identification Test
[3] https://www.kidscreen.org/espa%C3%B1ol/cuestionario-kidscreen/kidscreen-27/
[4] EDADES: Encuesta sobre alcohol y otras drogas en España
http://www.pnsd.msssi.gob.es/profesionales/sistemasInformacion/sistemaInformacion/encuestas_EDADES.htm

Following the line of researchers who have conducted studies through surveys, another problem that is often found is the collection of the same variables in different time periods. For example, if you want to analyze changes in behavior in a group of people over time, it is common to request that they repeat the same survey after a few months, so that, when dealing with this data, any other problems may be seen, together with the difficulty in processing the data and comparing the same population at different times [21].

*2.2. Semantic Technologies (Ontological Engineering and KBS) in SNA*

It is true that digital social networks are on the rise and more and more techniques related to artificial intelligence (data curation, data mining, natural language processing, etc.) are being applied [22–27]. However, in the field of social networking within the context that we have introduced in this article, there is not much scientific literature that has tried to solve or improve existing software systems by adding knowledge engineering.

Within the engineering of knowledge and the creation of conceptual models within the analysis of social networks, we find studies such as that of Khaled and colleagues [28], which created a recommendation system based on the semantic analysis of social networks in learning environments. Although the ontology they show in their research is interesting, the problem is that it focuses on a very particular problem that cannot be extrapolated, for example, in case studies such as the one explained in this article.

Another study combining semantic technologies, on this occasion, the use of RDF applied to SNA is that set out by researchers Raji and Surendran [29]. This manuscript discusses the importance of the use of knowledge engineering in social network analysis and mentions an applicable RDF model. However, this model can be used exclusively for data obtained through the extraction of data from online social networks. As already mentioned in this article, it cannot be applied to an SNA applied to data obtained through surveys.

Kassiri and Beloudha [30] propose a Unified Semantic Model (USM) based on the union of three standards combined with other three new ontologies to satisfy different needs found and related to the analysis of social networks applying semantic technologies. Its model is enriching from the point of view of digital social networks. However, they focus again on a specific case that cannot be extrapolated to our problem.

Guillaume Ereteo was the author who did the most work close to solving our problem [31–33]. The research of Guillaume Ereteo stands out when reviewing the existing scientific literature in the field of SNA and semantic technologies. Guillaume presents a system that carries out an SNA under a conceptual model generated by him and his research group [34]. This PhD thesis stems from the project called ISICIL Information Semantic Integration through Communities of Intelligence online.

The purpose of this research was to help analyze the characteristics of heterogeneous social networks that arise from the use of applications, applying semantic technologies to Social Network Analysis. SNA makes use of graphical algorithms and Semantic Web frameworks to represent and exchange knowledge through RDF, SPARQL and frameworks that use RDFS and OWL. There are therefore three main objectives:

- To transform data collected from social networks in the network to representations based on ontology.
- Carry out a Social Network Analysis that takes advantage of semantic technologies.
- Detect and semantically label communities of online social networks.

They use both OWL and the SPARQL query language in their methodology. They have developed the main ontology in their research which is based on FOAF (Friend Of A Friend) and an extension of it, in addition to the SIOC (Semantically-Interlinked Online Communities) and SCOT (Social Semantic Cloud of Tags) ontologies. FOAF is an ontology that describes people, their activities and their relationships with other people and objects. SIOC provides methods to interconnect different discussion sites, such as blogs, forums and mailing lists. Finally, SCOT represents a semantic model for social labeling.

To complete their ontology and the extraction of SNA, they use SPARQL through a proprietary tool called Corese that implements SPARQL with homomorphisms that facilitates the obtaining of RDF.

Their ontology called SemSNA allows a more complex analysis to be carried out using semantic technologies. Furthermore, their system can be used to improve and obtain more information in the future from an SNA than by applying any simple SNA application, thanks to semantics.

However, this ontology is only the beginning of an SNA. More concepts and properties are necessary to carry out a complete SNA.

**3. Material and methods:**

This research uses ontological engineering for knowledge modeling applied to a particular domain. An ontological knowledge model is an abstract structure of concepts, in which each concept has properties and relationships to represent the knowledge connotations. On the other hand, a knowledge model can thus be seen as a schema for knowledge description. When a model is implemented, the concepts are implemented into a domain ontology that allows the declarative knowledge and task ontology to be represented so that procedural knowledge can be inferred. The instances are then included into the schema to become a knowledge base. Making use of the different guidelines proposed in different studies [35–42], we propose four characteristics for the modeling of an OWL-based, Knowledge-Based System development as listed below:

1. *Analyzing the problem scenarios:* There are three kinds of knowledge: domain knowledge, task knowledge and inference knowledge. These correspond to an initial domain ontology, task ontology and semantic rules and SPARQL queries in a Knowledge-Based System. Thanks to the analysis of the problem scenario it is possible to identify the tasks and the task knowledge needed for problem solving in relation to the domain knowledge.

2. *Modeling the initial ontology:* This process of modeling the domain ontology may be achieved from different methodologies, for example, making use of Web sources such as open data and other published ontologies. It is possible to say that the domain ontology is hierarchical declarative knowledge. This task allows us to obtain a hierarchical taxonomy, different semantic elements such as properties (such as object properties, data properties, etc), classes, subclasses and instances that are contained in different classes that have been previously defined. Furthermore, it is necessary to use a specific vocabulary related to the domain of the problem.

3. *Modeling the rest of the ontology to solve the problem completely*: To carry out a task, it is necessary to create a task ontology in which the problem solving has been implemented. Making use of different kinds of properties, such as object properties, data properties and different characteristic of these, it is possible to infer other different properties that give rise to a different axiom for implementing the reasoning process.

4. *Developing the semantic rules required and SPARQL queries*: In a KBS, it is necessary to transfer the human expertise into machine understanding, therefore, this task is carried out by axioms. There are two kinds of axioms in KBS: logical axioms and non-logical axioms. The first type, logical axioms, are used to collect available known facts to infer implicit knowledge; nevertheless, non-logical axioms are formulas such as arithmetical operations. One way of programming non-logical axioms is to make use of semantic rules such as the SWRL or queries that modify the ontology into SPARQL Queries [35–41]. SWRL and SPARQL Queries are commonly used for building inference mechanism in OWL-based Knowledge-Based Systems.

The research process can be divided into four phases including: (1) searching for the ontologies phase, in which researchers search for different ontologies related to the domain of the problem to be solved; (2) the modeling phase, the domain ontology, task ontology and inference rules were jointly constructed in this phase; (3) the development phases, in which the team develop the Knowledge-Based System that makes use of the previously constructed ontology; and (4) experiment and evaluation phase, in which the system is tested using a real case and the results is compared with other validated applications to validate it. In Table 1 it is possible to see a summary of the different phases and tasks carried out.

**Table 1**
Research roles and responsibilities in each research phase.

| Phase | Role | | | |
|---|---|---|---|---|
| | Project leader | Domain experts | | Knowledge engineer |
| | | Expert in SNA | Nurse | |
| Initiation processes | Setup project meeting, research goals, and milestones. | Validate of SNA problem-solving method. | Analyze problem scenarios. | Creating axioms. |
| Modeling phase | Host modeling meetings to create domain ontology and task ontology. | Verify declarative knowledge and verification of problem solving. Help to identify tasks and subtasks. Help to identify the steps of semantic rules. | Suggest relevant declarative knowledge sources. | Define domain ontology construct and relevant instances. Define task ontology construct and properties. Define semantic rules and sparql queries through subtasks. |
| Development phase | Verify the KBS | Test inference mechanisms. | Test inference mechanisms. | Edit ontology and semantic rules using Protégé and edit SPARQL queries using Apache Fuseki. Integrate OWL-based KBS into web-based application. |
| Experimental phase | Verify experimental results | | Evaluating system performance using testing data. | Applying rules and queries to a real case and making a comparisonbetween KBS and UCINET, Cytoscape, Pajek and Gephi results. |

*3.1. Analyzing the problem scenarios*

The problem of carrying out a complete Social Network Analysis can be generally understood as the interaction between sources of knowledge related to people, other sources related to questionnaires (or surveys) and sources related to the terms of the Social Network Analysis. As shown in Fig. 1, multiple sources of knowledge are involved and there are also interactions between the sources of knowledge. An ontology model related to people is easier to conceptualize because there is an ontology called FOAF that has conceptualized concepts as a person and all his or her properties and the different kinds of relationships between people and other subjects [43]. In relation to the domain of questionnaires, as a concept it includes questions, different types of questions, answers, different types of answers, the answer to a question from a person and other concepts and their different relationships. The Social Network Analysis knowledge, for example, includes SNA concepts such as SNA indices such as degree, indegree, outdegree, betweenness and other relevant data of a network. Our solution needs to relate these different kinds of knowledge sources to offer us conclusions on a Social Network Analysis in a specific network of people who have previously completed a questionnaire.

To implement the analysis of the problem scenarios, two researchers specialized in Social Network Analysis have participated as consultants in verifying knowledge sources and problem scenarios from the beginning. After that, three knowledge engineers then formalized the problem solving in different non-logical axioms:

(1) Relating concepts between people and questionnaires

Thinking about a complete Social Network Analysis carried out by a Knowledge-Based System, firstly, a group of people needs to draw up a questionnaire with different kinds of questions. This questionnaire should be drawn up more than once because Knowledge-Based Systems need to compare results to reach conclusions on it. In these questionnaires there are different kinds of questions, answers and one kind of question is related to people who will complete this questionnaire. For example, we can find a question like: *How much time do you spend with any of the people of the list below?* and this question is answered by one group of people about this other group of people. So, this question provides information about the different relationships between a group of people. With this information it is possible to carry out a Social Network Analysis, obtaining different metrics values such as betweenness, centrality or degree of different people in a network and of a network itself.

To achieve formalized different concepts and to link all of the knowledge, firstly it is important to relate the concepts of the people and the questionnaires.

(2) Relating concepts between people and social networks

We should also point out the fact that the people's concepts need to be linked with the social network's concepts as well as those of the questionnaires. When a researcher specialized in Social Network Analysis creates a questionnaire, this questionnaire has questions to create one or more social network. So, it is known that people who complete this questionnaire will be a member of one or more networks. Knowing this, it is necessary to conceptualize the different links between the people and the networks.

People have different attributes within a network as well as different kinds of relationships and levels such as friendship, fellowship or family relationships for example.

(3) Relating concepts between questionnaires and social networks

Finally, the Knowledge-Based System needs to relate concepts on questionnaires and social networks. For example, a question like "*Indicate your gender*" provides information about a person who is member of a social network, specifically, this question provides the attribute of a person in a network.

*3.2. Modeling the initial ontology*

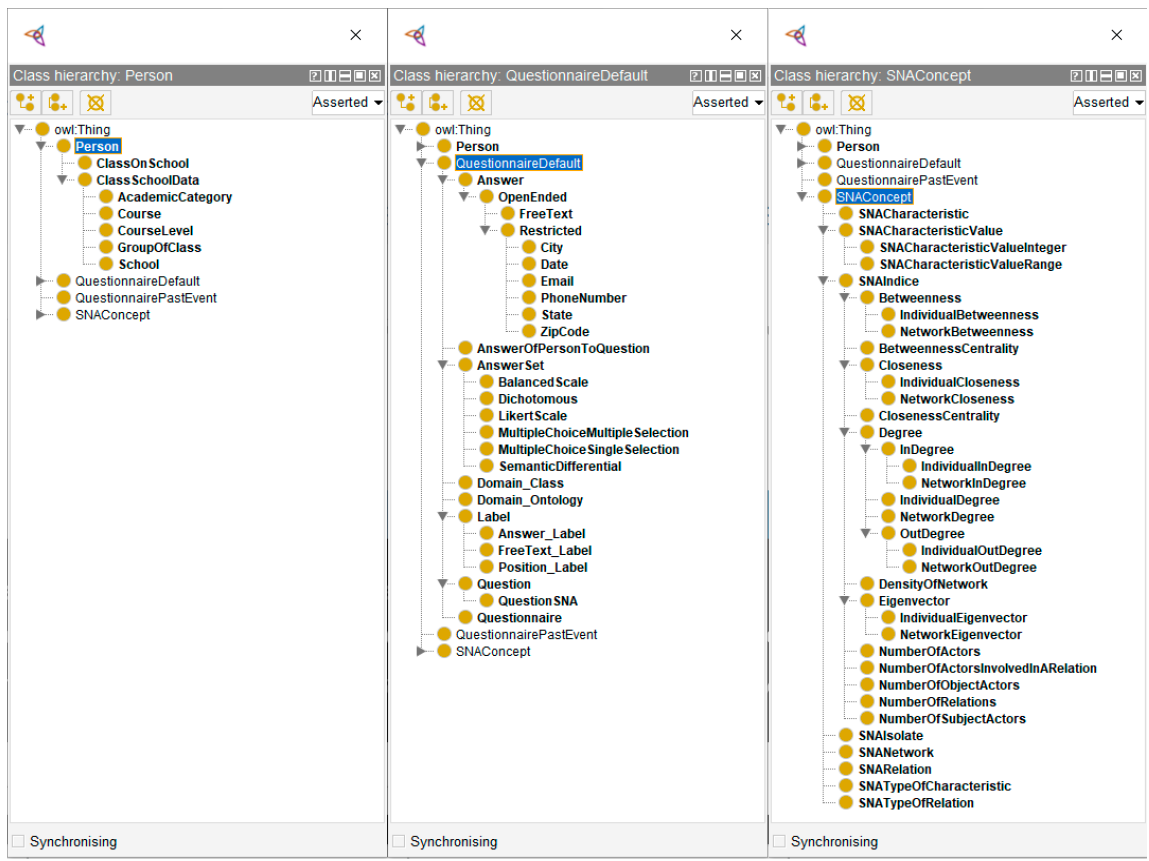

**Fig. 1.** Classes of OntoSNAQA

It should be noted that a domain ontology consists of a general conceptual structure and instances using an "is-a" to establish the subsumption relationships. By asserting instances it is possible to draw up the final concepts. For this reason, a domain ontology is a taxonomy that has not been created to solve problems, but is mainly about a common understanding of the domain for ontological sharing and reuse. Based on the analysis detailed in Section 3.1., before conceptualizing everything that is needed, it might be necessary to search for existing ontologies as explained in Section 2. In our problem, it was mainly necessary to conceptualize the three domains listed below:

- Domain of the people
- Domain of the questionnaires
- Domain of the Social Network Analysis

In order to transfer this knowledge into an OWL-based ontology called OntoSNAQA[5] (Ontology, Social Network Analysis, Questions & Answers), the Protégé is used to create the classes, properties and instances. Fig. 2 shows that the top-level concepts of the three different domains were mainly conceptualized. The child concepts and grandchild concepts are filled with constituent instances (see Fig. 1). Each instance can have its own internal property description (e.g., Fig. 2.). The concepts included are as follows:

(1) *Person:* This concept is completely reused from the FOAF ontology but it would be necessary to add new data properties and two news sub-concepts. Under each concept, the common terms are listed to provide reference and indexing for communication with other concepts and instances.

- ClassSchoolData: This concept establishes information on a class at a primary school because, in our study, we need this information about the people to create the different relationships automatically. Other sub-concepts such as *AcademicCategory, Course, CourseLevel, GroupOfClass* and *School* have been established in this concept.

- ClassOnSchool: This concept allows a complete class at a primary school to be represented. Instances of this class, can relate to different concepts about ClassSchoolData. We can see an example of an instance of this class in Fig. 2.

---

[5] Ontology is available at http://dx.doi.org/10.17632/gw8xmf74ws.1

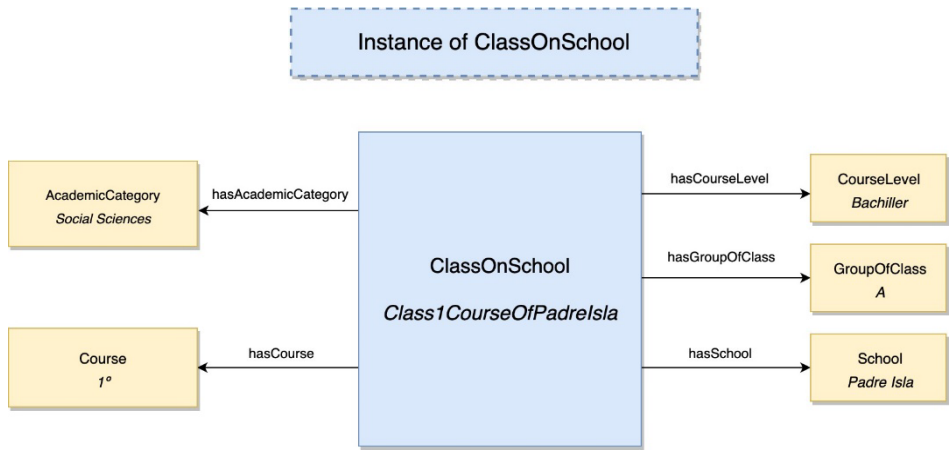

**Fig. 2.** Example of ClassOnSchool instance

(2) *QuestionnaireDefault:* This ontology is based in an ontology created by Alipour-Aghdam [44] but with a lot of new sub-concepts added. Furthermore, to achieve the objectives of this research, it was necessary to add three new concepts:

- QuestionSNA: It is a sub-concept of the *Question* class that helps us to represent questions related directly to the networks. For example, a question of this type could be: *How much time do you spend with any of the people on the list below?* This question will be repeated with all the members in a network. If the questionnaire is to be completed by 30 students in a class, this question will be asked 30 times to each student.

- QuestionnairePastEvent: This concept is necessary to represent a *Questionnaire* completed at a certain time. The questionnaires could be completed by users as many times as the researcher needs throughout his or her research. For example, a user can complete a questionnaire entitled *Breathalyzer risk questionnaire* in 2016 and he or she can do it again in 2017. Thanks to this, it is possible to relate to the different members of a network with the two events from the same questionnaire at different times.

- AnswerOfPersonToQuestion: This concept represents an answer of a user to a certain question in a certain questionnaire completed at a certain time. This concept allows the concepts of: *Person*, *QuestionnairePastEvent*, *Question* and *Answer* to be related.

We can see an example of different instances which interact with each other in the ontology developed in Fig 3.

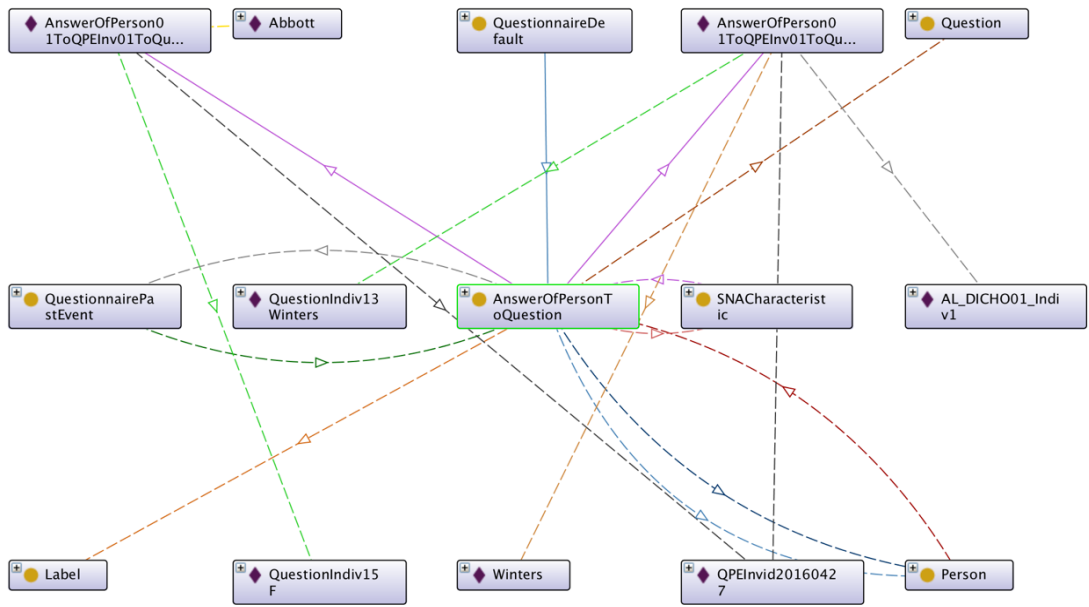

**Fig. 3.** Conceptual representation on a person who has answered a questionnaire at a certain time using OntoSNAQA.

(3) *SNAConcept:* As previously mentioned, within the scope of semantic technologies, the INRIA research group to which Dr. Guillaume Ereteo belongs, implemented a conceptual model that allowed the creation of instances within the domain of Social Network Analysis. This model had a series of classes through which typical concepts could be represented in an analysis of a social network such as: *SNAIndice, Degree, Betweenness, Closeness, etc*. However, the needs of the model presented, required a larger number of concepts to be added to make it more complete and meet the requirements of the initially proposed objectives. This is why it has been necessary to add 24 new classes to the initial model.

*3.3. Construction of task ontology*

A task ontology is used for enumerating and representing the specific problems to be solved. Besides constructing the conceptual structure, it is necessary to represent the different properties between the concepts in order to describe the problem-solving knowledge framework. OWL-based properties have different properties. The property values are first defined to separate asserted properties from inferred properties, both the known values and the unknown knowledge. The next step is to establish the corresponding domain and range of the properties. There are two types of properties, data properties and object properties. Data properties represent properties that use a basic data type in its range and object properties use instances of other classes in the range.

The task ontology has three main concepts related to the three domains that would be necessary to represent the achievement of the different objectives in this research. The domains of the different properties are mainly "Personal profile", "Questionnaire" and "Social Network Analysis" described below:

(1) *Personal profile*: There are 25 properties which have been designed for relating the domain of Person with concepts in the Questionnaires and Social Network Analysis. One asserted property is to give a name to the different concepts and twenty-four inferred properties to represent the different relationships between the Person and the Questionnaires and the Person and the Social Network Analysis. Thanks to these properties it is possible to establish that a person has answered a question in a questionnaire on a certain date, for example:

| Abott | *hasAnsweredToQuestionnairePastEvent* | *Quest01of20170530* |
|-------|---------------------------------------|---------------------|
|       | *hasAnsweredToQuestion*               | *QuestionGender*    |

It is also possible to represent the different results of a person within a network, for example:

| Abott | *isMemberOf* | *FriendshipRelation* |
|-------|--------------|----------------------|
|       | *hasAnswerOfPersonToQuestion* | *Answer01OfQPE01toQ01* |
|       | *hasIndividualDegree* | *DegreeOfAbottInFriendshipRelation* |
|       | *isStudentOf* | *Class1CourseOfPadreIsla* |
|       | *hasRelation* | *AbottFriendOfClarck* |
|       | *hasCharacteristicOfPerson* | *MaleGender* |

(2) *Questionnaire:* Forty-one properties are designed, including ten asserted properties designed to annotate the date on which a person completed a questionnaire, the value of a label which represents an answer, a text of a question and other concepts. Inferred properties allow Questionnaire concepts to be related with Person concepts and Social Network Analysis Concepts. For example, we can see below the representation of a question answered by a person in a questionnaire on a certain date in which the person has answered "male" to the question "What is your gender?". Furthermore, this question represents a characteristic of a person in a network:

(3)

| | | |
|---|---|---|
| *QuestionGender* | has_Question_Text | "What is your gender?" |
| | hasAnswer | AnswerOfGender |
| | isQuestionOf | QPE01 |
| | isQuestionOfCharacteristicOfPerson | MaleGender |
| *MaleGender* | isAnsweredBy | Abott |
| | hasLabel | MaleLabel |
| | isAnswerOfPersonToQuestionOf | Answer01OfQPE01toQ01 |
| *MaleLabel* | has_Value | "Male" |

(4) *Social Network Analysis:* There are sixty-five properties needed to represent all of the relationships between the Social Network Analysis and the other two domains. Five properties are asserted and these properties have been used to represent the value or name of a characteristic in the network, the name of a network or the date of creation of a network from a Social Network Analysis. So, there are sixty properties created to represent the relationships between the Social Network Analysis concepts and the concepts of the Person domain and the Questionnaire domain. An example of a network related to a questionnaire and that related to a person is represented below:

| | | |
|---|---|---|
| *FriendshipNetwork* | has_Network_Name | "Friendship relation" |
| | hasMember | Abott |
| | isNetworkOfQPE | QPE01 |
| | isNetworkOfTypeOfRelation | FriendshipRelation |

**Table 2**
Task ontology

| Mainly concept | Class | Attribute definition | | | Rules & queries |
|---|---|---|---|---|---|
| | | Attribute name | Type | Range | |
| Personal profile | Person | hasAnsweredToQuestion | Inferred | Question | Rule-1 |
| | | hasAnsweredToQuestionnairePastEvent | Inferred | AnswerOfPersonToQuestion | |
| | | hasCharacteristicOfPerson | Inferred | SNACharacteristic | |
| | | hasIndividualBetweenness | Inferred | IndividualBetweenness | |
| | | hasIndividualCloseness | Inferred | IndividualCloseness | |
| | | hasIndividualDegree | Inferred | IndividualDegree | |
| | | hasIndividualEigenvector | Inferred | IndividualEigenvector | |
| | | hasIndividualInDegree | Inferred | IndividualInDegree | |
| | | hasIndividualOutDegree | Inferred | IndividualOutDegree | |
| | | hsIsolateInstanceOfPerson | Inferred | SNAIsolate | |
| | | hasRelation | Inferred | SNARelation | |
| | | isMemberOf | Inferred | SNANetwork | |
| | | isStudentOf | Inferred | ClassOnSchool | |
| | | isTargetOfRelation | Inferred | SNARelation | |
| | ClassSchoolData | has_Data_Name | Asserted | (string) | |
| | AcademicCategory | isAcademicCategoryOf | Inferred | ClassOnSchool | |
| | Course | isCourseOf | Inferred | ClassOnSchool | |
| | CourseLevel | isCourseLevelOf | Inferred | ClassOnSchool | |
| | GroupOfClass | isGroupOfClassOf | Inferred | ClassOnSchool | |
| | School | isSchoolOf | Inferred | ClassOnSchool | |
| | ClassOnSchool | hasAcademicCategory | Inferred | AcademicCategory | |
| | | hasCourse | Inferred | Course | |
| | | hasCourseLevel | Inferred | CourseLevel | |
| | | hasSchool | Inferred | School | |
| | | hasStudent | Inferred | Person | |

| Questionnaire | QuestionnairePastEvent | has_Date_Start | Asserted | (datetime) |
| --- | --- | --- | --- | --- |
| | | has_Date_End | Asserted | (datetime) |
| | | has_Event_Id | Asserted | (int) |
| | | hasAnsweredToQuestionnairePastEvent | Inferred | Person |
| | | hasAnswerOfPersonToQuestion | Inferred | AnswerOfPersonToQuestion |
| | | hasCharacteristicOfQPE | Inferred | SNACharacteristic |
| | | hasNetwork | Inferred | SNANetwork |
| | | hasQuestionnaire | Inferred | Questionnaire |
| | | isAnsweredByPerson | Inferred | Person |
| | Answer | has_Number_Of_Answer_Label | Asserted | (int) |
| | | hasLabel | Inferred | Label |
| | | isAnswerOf | Inferred | AnswerSet |
| | AnswerOfPersonToQuestion | hasAnswered | Inferred | Label |
| | | hasAnsweredTo | Inferred | Question |
| | | isAnAnsweringRelatingTo | Inferred | Person |
| | | isAnswerOfPersonOfCharacteristic | Inferred | SNACharacteristic |
| | | isAnswerOfPersonToQuestionOf | Inferred | Person |
| | | isAnswerOfQuestionnairePastEvent | Inferred | QuestionnairePastEvent |
| | AnswerSet | has_Number_Of_Answers | Asserted | (int) |
| | | has_Number_Of_Selected | Asserted | (int) |
| | | hasAnswer | Inferred | Answer |
| | | isAnswerSetOf | Inferred | Question |
| | Domain_Class | isDomainClassOf | Inferred | Domain_Ontology |
| | Domain_Ontology | has_Domain_IRI | Asserted | (string) |
| | | has_Domain_Name | Asserted | (string) |
| | | hasDomainClass | Inferred | Domain_Class |
| | | isDomainOntologyOf | Inferred | Question |
| | Answer_Label | has_Value | Asserted | (string) |
| | | isAnswerOfCharacteristicValue | Inferred | SNACharacteristicValue |

| | | | | | |
|---|---|---|---|---|---|
| Social Network Analysis | | isAnswerOfTypeOfRelation | Inferred | SNATypeOfRelation | |
| | | isLabelOf | Inferred | Answer | |
| | Question | has_Question_Text | Asserted | (string) | |
| | | hasAnswerSet | Inferred | AnswerSet | |
| | | hasDomainOntology | Inferred | Domain_Ontology | |
| | | hasMergedDomainClass | Inferred | Domain_Class | |
| | | isQuestionOf | Inferred | Questionnaire | |
| | | isQuestionOfCharacteristicOfPerson | Inferred | SNACharacteristic | |
| | | isQuestionOfTypeOfCharacteristic | Inferred | SNATypeOfCharacteristic | |
| | | isQuestionOfTypeOfRelation | Inferred | SNATypeOfRelation | |
| | Questionnaire | hasQuestion | Inferred | Question | |
| | | isQuestionaireOf | Inferred | QuestionnairePastEvent | |
| | SNACharacteristic | has_Characteristic_Name | Asserted | (string) | |
| | | has_Characteristic_Value | Asserted | (string) | |
| | | hasCharacteristicValue | Inferred | SNACharacteristicValue | |
| | | isAnCharacteristicOfPersonOfQuestion | Inferred | Question | |
| | | isCharacteristicOfAnswerOfPerson | Inferred | AnswerOfPersonToQuestion | Rule-5 |
| | | isCharacteristicOfNetwork | Inferred | SNANetwork | Rule-3 |
| | | isCharacteristicOfPerson | Inferred | Person | Rule-3 and 4 |
| | | isCharacteristicOfQPE | Inferred | QuestionnairePastEvent | Rule-3 and 4 |
| | | isCharacteristicOfType | Inferred | SNATypeOfCharacteristic | Rule-3 and 4 |
| | SNACharacteristicValue | isCharacteristicValueOf | Inferred | SNACharacteristic | Rule-3 |
| | | isCharacteristicValueOfAnswer | Inferred | Answer_Label | |
| | | isPossibleCharacteristicValueOf | Inferred | SNATypeOfCharacteristic | |
| | SNAIndice | has_SNA_Value | Asserted | (float) | Query-1 to 11 |
| | IndividualBetweenness | isIndividiualBetweennessOfPerson | Inferred | Person | Rule-6 |
| | NetworkBetweenness | isNetworkBetweennessOfNetwork | Inferred | SNANetwork | Rule-7 |
| | IndividualCloseness | isIndividualClosenessOfPerson | Inferred | Person | Rule-6 |
| | NetworkCloseness | isNetworkClosenessOfNetwork | Inferred | SNANetwork | Rule-7 |
| | IndividualInDegree | isIndividualInDegreeOfPerson | Inferred | Person | Rule-6 |

| | | | | |
|---|---|---|---|---|
| NetworkInDegree | isNetworkInDegreeOfNetwork | Inferred | SNANetwork | Rule-7 |
| IndividualDegree | isIndividualDegreeOfPerson | Inferred | Person | Rule-6 |
| NetworkDegree | isNetworkDegreeOfNetwork | Inferred | SNANetwork | Rule-7 |
| IndividualOutDegree | isIndividualOutDegreeOfPerson | Inferred | Person | Rule-6 |
| NetworkOutDegree | isNetworkOutDegreeOfNetwoerk | Inferred | SNANetwork | Rule-7 |
| DensityOfNetwork | isDensityOfNetworkOf | Inferred | SNANetwork | Rule-7 |
| IndividualEigenvector | isIndividualEigenvectorOf | Inferred | Person | Rule-6 |
| NetworkEigenvector | isNetworkEigenvectorOf | Inferred | SNANetwork | Rule-7 |
| NumberOfActors | isNumberOfActorsOf | Inferred | SNANetwork | Rule-7 |
| NumberOfActorsInvolvedInARelation | isNumberOfActorsInvolvedInARelationOf | Inferred | SNANetwork | Rule-7 |
| NumerOfObjectActors | isNumberOfObjectActorsOf | Inferred | SNANetwork | Rule-7 |
| NumberOfRelations | isNumberOfRelationsOf | Inferred | SNANetwork | Rule-7 |
| NumberOfSubjectActors | isNumberOfSubjectActorsOf | Inferred | SNANetwork | Rule-7 |
| SNAIsolate | isIsolateInstanceOfNetwork | Inferred | SNANetwork | |
| | isIsolateInstanceOfPerson | Inferred | Person | |
| SNANetwork | has_Date | Asserted | (datetime) | |
| | has_Network_Name | Asserted | (string) | |
| | hasCharacteristicOfNetwork | Inferred | SNACharacteristic | |
| | hasDensityOfNetwork | Inferred | DensityOfNetwork | |
| | hasIsolateInstanceOfNetwork | Inferred | SNAIsolate | |
| | hasMember | Inferred | Person | Rule-1 |
| | hasNetworkBetweenness | Inferred | NetworkBetweenness | |
| | hasNetworCloseness | Inferred | NetworkCloseness | |
| | hasNetworkDegree | Inferred | NetworkDegree | |
| | hasNetworkEigenvector | Inferred | NetworkEigenvector | |
| | hasNetworkInDegree | Inferred | NetworkInDegree | |
| | hasNetworOutDegree | Inferred | NetworkOutDegree | |
| | hasNumberOfActors | Inferred | NumberOfActors | |
| | hasNumberOfActorsInvolvedInARelation | Inferred | NumberOfActorsInvolvedInARelation | |
| | hasNumberOfObjectActors | Inferred | NumberOfObjectActors | |
| | hasNumberOfRelations | Inferred | NumberOfRelations | |
| | hasNumberOfSubjectActors | Inferred | NumberOfSubjectActors | |
| | hasRelationOfPerson | Inferred | SNARelation | |
| | isNetworkOfQPE | Inferred | QuestionnairePastEvent | |
| | isNetworkOfTypeOfRelation | Inferred | SNATypteOfRelation | |
| SNARelation | isRelationOfNetwork | Inferred | SNANetwork | Rule-2 |
| | isRelationOfPerson | Inferred | Person | Rule-2 |
| | isRelationOfType | Inferred | SNATypeOfRelation | Rule-2 |
| | isRelationWith | Inferred | Person | Rule-2 |
| | isRelationOfQpe | Inferred | QuestionnairePastEvent | Rule-2 |
| SNATypeOfCharacteristic | hasPossibleCharacteristicValue | Inferred | SNACharacteristicValue | |
| | isCharacteristicOfQuestion | Inferred | Question | |
| | isTypeOfCharacteristicOf | Inferred | SNACharacteristic | |
| SNATypeOfRelation | hasAnswerOfTypeOfRelation | Inferred | Answer_Label | |
| | hasQuestionOfTypeOfRelation | Inferred | Question | |
| | isTypeOfRleationOfNetwork | Inferred | SNANetwork | |
| | isTypeOfRelationOf | Inferred | SNARelation | |

*3.4. Development of semantic rules:*

In the task ontology detailed in Table 2, there are 32 inferred properties which are designed as sub-tasks. These properties require semantic rules in SWRL and SPARQL queries to combine the related facts for inference. We collect the practical problem solving experiences before creating semantic rules, and then use the "premise -> conclusion" logic form to describe the solving process [45]. The rules start with the concept to which the property belongs, and then links the concept to other facts in a step-by-step manner until the objective is achieved. Each step is expressed as an atom and the rule is expressed in the form of "(atom$_1$^...^atom$_n$) -> Consequence" to express the cause-effect relationship.

In this study, seven SWRL-based rules and eleven SPARQL queries have been developed to obtain the concepts of the "Personal profile", "Questionnaire" and "Social Network Analysis".

(1) Rule-1: Linking a person with a network and a questionnaire on a certain date:

The first rule (Rule-1) represents the link between three concepts: *Person*, *SNANetwork* and *AnswerOfPersonToQuestionOf*.

| | |
|---|---|
| Person(?p) ^ SNANetwork(?net) ^ AnswerOfPersonToQuestion(?aoptq) ^ isAnswerOfQuestionnairePastEvent(?aoptq, ?qpe) ^ hasNetwork(?qpe, ?net) -> hasMember(?net, ?p) ^ **hasAnsweredToQuestionnairePastEvent(?p, ?qpe)** | **Rule-1** |

(2) Rule-2: A rule used to create instances of relationships between different people in a network

The Rule-2 searches to see whether an answer is of type "Answer Of Type Of Relation" (*isAnswerOfTypeOfRelation*) and it adds a relationship with the corresponding person of the corresponding type. For example, the answer "I'm always with Juan" answered by Laura implied that:

- I am always with => it is an answer to a relationship of friendship, partnership and acquaintance.

- So, in this case, three instances of class SNARelation will be created between Laura and Juan. These instances represent the three relationships between both.

This creation of a new instance is possible thanks to *swrlx:makeOWLThing,* belonging to the *ExtensionsBuiltInLibrary*. Once the instance is generated, initially in class *Thing*, in the consequent class, it is indicated that the instance is of type *SNARelation*.

| | |
|---|---|
| isAnswerOfTypeOfRelation(?al, ?tor) ^ isTypeOfRelationOfNetwork(?tor, ?net) ^ swrlx:makeOWLThing(?rel, ?p, ?q) ^ isAnswerOfPersonToQuestionOf(?a, ?p) ^ Person(?q) ^ isAnswerOfQuestionnairePastEvent(?a, ?qpe) ^ Person(?p) ^ hasAnswered(?a, ?al) ^ SNANetwork(?net) ^ isNetworkOfQPE(?net, ?qpe) ^ isAnAnswerRelatingTo(?a, ?q) ^ differentFrom(?p,?q) -> **isRelationOfType(?rel, ?tor) ^ isRelationOfNetwork(?rel, ?net) ^ isRelationWith(?rel, ?q) ^ SNARelation(?rel) ^ isRelationOfQPE(?rel, ?qpe) ^ isRelationOfPerson(?rel, ?p)** | **Rule-2** |

(3) Rule-3: this rule assigns a characteristic to a person within a network in relation to his or her answer to a questionnaire.

Some answers give value to a characteristic of a member within a network. This rule analyses whether the answer gives value to a characteristic of a person in a network and, if so, this rule generates a new instance of type characteristic, giving it the properties necessary to relate a person in a network with a type of characteristic.

| | |
|---|---|
| Person(?p) ^ hasAnsweredToQuestion(?p, ?atoq) ^ hasAnswered(?atoq, ?al) ^ isAnswerOfCharacteristicValue(?al, ?av) ^ isPossibleCharacteristicValueOf(?av, ?atype) ^ isAnswerOfQuestionnairePastEvent(?a, ?qpe) ^ hasNetwork(?qpe, ?net) ^ swrlx:makeOWLThing(?aop, ?av, ?p) -> **SNACharacteristic(?aop) ^ isCharacteristicOfNetwork(?aop, ?net) ^ isCharacteristicOfPerson(?aop, ?p) ^ isCharacteristicOfQPE(?aop, ?qpe) ^ isCharacteristicValueOf(?av, ?aop) ^ isCharacteristicOfType(?aop, ?atype)** | **Rule-3** |

(4) Rule-4: this rule creates an instance of class SNACharacteristic. This instance is a characteristic of a person that corresponds to the answer of this person to a question in a questionnaire answered at a certain time so later, through a SPARQL query, it will be possible to add the value corresponding to the sum of these answers, that it to say:

- If Juan is member of a network *Friendship* and there is a characteristic, for example *AuditValue*, that has an integer value, a new instance will be generated.

- This instance *AuditOfJuanInFriendship* has a relationship with an instance of an SNACharacteristic and this characteristic belongs to a type of characteristic, to a person, to a network and to a questionnaire at a certain time.

Person(?p) ^ SNACharacteristicValueInteger(?av) ^ isPossibleCharacteristicValueOf(?av, ?toa) ^ hasMember(?net, ?p) ^ hasAnsweredToQuestionnairePastEvent(?p, ?qpe) ^ swrlx:makeOWLThing(?aop, ?p, ?av) -> SNACharacteristic(?aop) ^ isCharacteristicOfType(?aop, ?toa) ^ isCharacteristicOfPerson(?aop, ?p) ^ isCharacteristicOfNetwork(?aop, ?net) ^ isCharacteristicOfQPE(?aop, ?qpe) **Rule-4**

(5) Rule-5: after Rule-3 has been applied, thanks to this new rule, it is possible to find out which answer of a person is related to this new characteristic, for example:

- Juan has a characteristic *Gender* in a network called *Friendship*.

- This network *Friendship* is related to questionnaire answered at a certain time called *QPE01*.

- So, there is an answer of Juan in questionnaire *QPE01* which is related to the characteristic *Gender*.

SNACharacteristic(?aop) ^ isCharacteristicOfPerson(?aop, ?p) ^ isCharacteristicOfType(?aop, ?toa) ^ SNACharacteristicValueInteger(?av) ^ isPossibleCharacteristicValueOf(?av, ?toa) ^ isCharacteristicOfQuestion(?toa, ?q) ^ isAnswerOfPersonToQuestionOf(?aoptq, ?p) ^ hasAnsweredTo(?aoptq, ?q) -> **isCharacteristicOfAnswerOfPerson(?aop, ?aoptq)** **Rule-5**

(6) Rule-6: this rule creates the SNAConcepts of a person in a network. Thanks to this rule it is possible to create all instances relating a person with a network in terms of Social Network Analysis, concepts such as: *IndividualBetweenness, IndividualCloseness, IndividualDegree*, etc.

Person(?p) ^ SNANetwork(?net) ^ hasMember(?net, ?p) ^ **Rule-6**

swrlx:makeOWLThing(?bw, ?p, ?net) ^ swrlx:makeOWLThing(?cn, ?p, ?net) ^ swrlx:makeOWLThing(?dg, ?p, ?net) ^ swrlx:makeOWLThing(?in, ?p, ?net) ^ swrlx:makeOWLThing(?ou, ?p, ?net) ^ swrlx:makeOWLThing(?ei, ?p, ?net) -> **IndividualBetweenness(?bw)** ^ **isIndividualBetweennessOfPerson(?bw, ?p)** ^ **isIndividualBetweennessOfNetwork(?bw, ?net)** ^ **IndividualCloseness(?cn)** ^ **isIndividualClosenessOfPerson(?cn, ?p)** ^ **isIndividualClosenessOfNetwork(?cn, ?net)** ^ **IndividualDegree(?dg)** ^ **isIndividualDegreeOfPerson(?dg, ?p)** ^ **isIndividualDegreeOfNetwork(?dg, ?net)** ^ **IndividualInDegree(?in)** ^ **isIndividualInDegreeOfPerson(?in, ?p)** ^ **isIndividualInDegreeOfNetwork(?in, ?net)** ^ **IndividualOutDegree(?ou)** ^ **isIndividualOutDegreeOfPerson(?ou, ?p)** ^ **isIndividualOutDegreeOfNetwork(?ou, ?net)** ^ **IndividualEigenvector(?ei)** ^ **isIndividualEigenvectorOfPerson(?ei, ?p)** ^ **isIndividualEigenvectorOfNetwork(?ei, ?net)**

(7) Rule-7: this rule creates the SNAConcepts of a network. Thanks to this rule it is possible to create all instances relating a network to the different terms of Social Network Analysis, concepts such as: *NetworkBetweenness, NetworkDegree, NetworkInDegree* etc.

| | |
|---|---|
| SNANetwork(?net) ^ isNetworkOfQPE(?net, ?p) ^ swrlx:makeOWLThing(?bw, ?p, ?net) ^ swrlx:makeOWLThing(?cn, ?p, ?net) ^ swrlx:makeOWLThing(?dg, ?p, ?net) ^ swrlx:makeOWLThing(?in, ?p, ?net) ^ swrlx:makeOWLThing(?ou, ?p, ?net) ^ swrlx:makeOWLThing(?ei, ?p, ?net) ^ swrlx:makeOWLThing(?noa, ?p, ?net) ^ swrlx:makeOWLThing(?noair, ?p, ?net) ^ swrlx:makeOWLThing(?nooa, ?p, ?net) ^ swrlx:makeOWLThing(?nosa, ?p, ?net) ^ swrlx:makeOWLThing(?nor, ?p, ?net) ^ swrlx:makeOWLThing(?don, ?p, ?net) -> **NetworkBetweenness(?bw)** ^ **hasNetworkBetweenness(?net, ?bw)** ^ **NetworkCloseness(?cn)** ^ **hasNetworkCloseness(?net, ?cn)** ^ **NetworkDegree(?dg)** ^ **hasNetworkDegree(?net, ?dg)** ^ **NetworkInDegree(?in)** ^ **hasNetworkInDegree(?net, ?in)** ^ **NetworkOutDegree(?ou)** ^ **hasNetworkOutDegree(?net, ?ou)** ^ **NetworkEigenvector(?ei)** ^ **hasNetworkEigenvector(?net, ?ei)** ^ **hasNumberOfActors(?net, ?noa)** ^ **hasNumberOfSubjectActors(?net, ?nosa)** ^ **hasNumberOfObjectActors(?net, ?nooa)** ^ **hasNumberOfRelations(?net, ?nor)** ^ **hasNumberOfActorsInvolvedInARelation(?net, ?noair)** ^ **hasDensityOfNetwork(?net, ?don)** | **Rule-7** |

Once the SWRL-based rules have been implemented in the ontology and applied to a particular case, this would become an ontology in which there would be instances that represent the values of a Social Network Analysis, such as:

Intermediation level: represented by the NetworkBetweenness and IndividualBetweenness classes.

- Sum of all relationships in a network: represented by class *NumberOfRelations*.
- Sum of all members: represented by the *NumberOfActors, NumberOfObjectActors, NumberOfSubjectActors, NumberOfActorsInvolvedInARelation*.

- Density of a network: represented by the class *DensityOfNetwork*.
- Sum of all isolated members: represented by the class *SNAIsolate*.
- Closeness level: represented by the *NetworkCloseness* and I*ndividualCloseness* classes.
- Degree of entry: represented by the *NetworkInDegree* and *IndividualInDegree* classes.
- Degree of output: represented by the *NetworkOutDegree* and *IndividualOutDegree* classes.
- Complete degree: represented by the *NetworkDegree* and I*ndividualDegreee* classes.
- Own vector: represented by the *NetworkEigenvector* and *IndividualEigenvector* classes.

Each of the instances will have a numerical value, represented by the data property *has_SNA_Value*. To make these calculations, it has been necessary to generate a series of queries in SPARQL Update, which can be processed later with environments such as Apache Fuseki Server, Stardog or Virtuoso. All of the queries are described in Appendix A and these are available at http://dx.doi.org/10.21227/z6pj-je73 .

## 4. Experimental validation and case study:

*4.1. Introduction to a case study*

The line of work that gives rise to this research is based on the fact that alcohol consumption among adolescents is a social and public health problem [6]. Research so far has found that contacts between adolescents can influence the habit of consumption. Specifically, the adolescent's social network includes diverse contact patterns in the classroom, which pose a risk in the consumption of drugs, such as the negative influence of peers, social norms that govern relationship groups, or the position they occupy within the social network. However, there is a lack of studies that look into how educational leaders and other multidisciplinary collectives perceive these contacts, in order to obtain useful information to transfer ethnographic information to the quantitative analysis of the SNA and design strategies in which the stakeholders are incorporated.

The objective of the case study was to analyze the perspective of teachers as to the position and pattern of behavior that adolescents occupy in the classroom and their relationship with the high risk of alcohol intake.

In this research, a study based on mixed methods was carried out in two schools of Compulsory Secondary Education, using random sampling according to inclusion criteria, with a total of 10 teachers. The methodology used was quantitative for the analysis of the structural data of the networks of contacts through an Analysis of Social Networks (self-administered questionnaire) and qualitative, through discussion groups, with the education officer responsible for the classrooms.

After obtaining the data through the questionnaires, the next step is to analyze the "raw" data to later reach a series of conclusions. The justification of this research begins in this step of the project.

All of the ontology explained throughout this document, as well as the SWRL rules and SPARQL queries, can and has helped to carry out this research, saving time for the health expert in drawing up the questions and allowing the values and conclusions to be obtained with a reduced error as it is an automated process, eliminating human error in the reading and transcription of data, among others. Another important feature is that it is now possible to visualize and obtain a result from the social network analysis.

*4.2. Web-based application development using the ontology*

For the primary purpose of testing the conceptual model generated in this research, a web application has been made that makes use of it. This web solution is mainly aimed at research experts, mainly from the health field, who wish to carry out studies on a population group based on personalized surveys. This application will allow the user to carry out the following functions:

- To create personalized surveys to send to a certain group of people.
- To carry out the Social Network Analysis on the data obtained through the surveys of the different users.

- To obtain an approximate interpretation of the data obtained in the Social Network Analysis carried out.

The proposed application has a Login System, through which the researcher will access a personalized control panel in which he or she will be able to carry out the necessary tasks for his or her research. This Login system has been made in the PHP programming language, using programming oriented objects, MySQL database and HTML5, Javascript (jQuery, AJAX) and CSS3 has been made use of for the web interface. It is also worth mentioning the use of the Apache Jena Fuseki Server to make direct queries to the ontology, as well as its repopulation from the administrator and manager environment.

Furthermore, it has been necessary to use specific libraries to manage the creation of different graphs, specifically the SigmaJS library (http://sigmajs.org/) used in other studies of interest [46], as well as its different plugins, highlighting especially the Louvain statistics manager, based on this algorithm [47].

The application initially has three roles:

- Super administrator: they have full permission to manage validated questionnaires, as well as users, pollsters and respondents.
- Surveyor: they can add validated questionnaires, under review by a super administrator, and you can only access your questions, surveys and interviewers.
- Surveyed: they only have access to the platform to carry out the questionnaires that their interviewer needs.

The interface is friendly and simple, without needing to have a high level of computer knowledge at the programming level. We came to this conclusion after obtaining a positive response from fourteen researchers from different scientific branches, all different from engineering: nursing, psychology, anthropology and pharmacy.

With this application, the researcher can carry out the following tasks:

- Manage validated questionnaires
- Manage questions for questionnaires

- Manage questionnaires
- Manage pollsters
- Analyze data from surveys conducted

The respondents will only access a control panel where they can see the surveys that have been assigned to them to answer.

This application has the following functionalities, based on the type of information that can be obtained from the data from the questionnaires that have been inserted into it in this research and the aforementioned requirements.

- Load the set of data from the answers to the questionnaire and show the different networks that can be useful to find out the habit of alcohol consumption in relation to friendship (at school level) and family networks.
- Show the data graphs and the results of the analysis on the social patterns of alcohol consumption taking into account the different levels of relationship with a partner (acquaintance, partner and friend) and also within the family environment, as well as basic information on the relationships with the individuals outside the school / class and the family circle.
- Show basic data and a description similar to a report on any individual in the network as to their alcohol consumption status, especially in relation to the risk of alcohol consumption disorder or any type of relevance within the network that the individual may experience (be it a mediator, an influencer for others, etc.). This report will be the result of different calculations carried out by SNA and other techniques.
- Show, for each individual, who can act as an influence on him / her.
- Show a description similar to the aforementioned report, but for social networks and entire groups that can be found.
- Show and study the level of alcohol consumption in relation to gender and the environment.

- Show and report the relationship that may exist between alcohol consumption, socioeconomic status, self-perception and self-efficacy.
- Show and report factors that may be related to the level of alcohol consumption, such as the use of politoxicomaniacs and the consumer and relationship environments.

*4.3. Description of validation method to the case study*

In the previous section, the web application that makes use of the ontology developed throughout this research has been explained [48]. However, what is really interesting and what determines whether the model works and that the data shown are valid, is the comparison of said data with an environment that has already been validated previously in the carrying out of the SNA.

For this reason, it was decided to carry out the social network analysis on the particular case study with the UCINET tool, Cytoscape, Pajek and Gephi, so that by making the aforementioned relevant SPARQL queries, it is possible to make a comparison of whether the results are exactly the same or if there are discrepancies in the system that has been generated.

Next, we will describe the data obtained using the UCINET tool in the particular case that arose in this research, as well as the data obtained through the ontology generated and populated with the data in question.

*4.4. Introduction to the social network analysis to be applied*

The particular case to which the generated conceptual model would be applied, as well as the aforementioned system, had as its main objective to analyze the perspective of the teaching staff as to the position and pattern of behavior that adolescents occupy in the classroom and their relationship with the high risk of alcohol consumption.

For this, a survey was carried out that included a total of 252 fixed generic questions and a variable number of questions related to social networks. This variability depended on the number of respondents, since there were two questions in this questionnaire that related the respondents to each other. If the questionnaire was completed by a group of forty people, for example, forty students in a school class, the total number of questions would be $252 + 80 =$

332. However, if the same questionnaire had been completed by a group of eighty people, there would have been the 252 common questions plus the 2 questions related to the eighty people, which would make a total of 252 + 160 = 412 questions in total.

This questionnaire was carried out for the first time by different students aged 16 and 17 in 9 different classes, with a total of 214 people surveyed, with a total of 14,555 answered questions to be analyzed. These 214 people represent several classrooms from different schools. To carry out our study we have selected a classroom of 38 people. The first step in the research was carried out manually, that is, the research group asked the different students to answer a survey using an online form.

This form stored the answers on a spreadsheet. This questionnaire had different standardized forms, specifically the following: AUDIT, FAS II, KIDSCREEN 27, STUDES.

The research group, once the data was obtained, took an average of 15 minutes per respondent to obtain the sum of the individual values of the questions corresponding to these forms.

Subsequently, they used Social Data Analysis tools for specific data blocks, including: GEPHI [49], Pajek [50] and UCINET [10]. With this methodology, an average of two hours was necessary to be able to obtain graphs in each of the 9 classes surveyed, having to analyze the meaning of the different data later.

In a second phase, they returned to carry out the data collection from the aforementioned questionnaire with the tool developed and explained in this article. Although the creation of the survey took them a total of 14 hours, including the involvement of the 38 users who had to carry out the survey, once all the users filled out their respective questionnaires, they obtained the results class by class immediately.

Next, we will detail the results of the analysis of a social network belonging to one of the nine classes at the school through UCINET, Cytoscape, Pajek, Gephi and through the application developed in this research that makes use of the generated ontology.

## 5. Results:

As mentioned above, the model, through the use of different SPARQL queries, is able to obtain different significant measurements when carrying out a personal social network analysis. That is why the results obtained through the different SNA programs (UCINET, Cytoscape, Pajek and Gephi) that determine the following values of a social network analysis are shown below:

Measurements from each network:

- Total number of relationships in a network
- Total number of stakeholders within a network
- Total number of members subject within a network
- Total number of total members within a network
- Total number of members that have at least one relationship within a network
- Average degree of a network
- Density of a network
- Total number of nodes without connection within a network

Measures obtained for each actor within the network:

- Degree of each actor within a network
- Indegree of each actor within a network
- Outdegree of each actor within a network

Within the study carried out by the population group made up of thirty-eight students aged 16 and 17 from a school, three networks have been differentiated according to the following criteria:

- One of the questions included in the questionnaire was responsible for how much time the respondent and his classmates spent together.
- There were five possible answers: never, almost never, sometimes, almost always, always.

- Each response had a numerical value of between 1 and 5 according to the previously expressed request and that can be seen in the table shown below in Table 3:

**Table 3**

List of labels with the value in a questionnaire.

| Answer | Value |
|---|---|
| Never | 1 |
| Almost never | 2 |
| Sometimes | 3 |
| Almost always | 4 |
| Always | 5 |

- Based on this weighting, three types of relationships have been determined. Table 4 shows these relationships.

**Table 4**

Values needed to consider a relation between two people in a network.

| Relationship | Value |
|---|---|
| Friendship | 4 and 5 |
| Workmate | 3,4 and 5 |
| Acquaintance | 2,3,4 and 5 |

Once the three networks chosen for the validation of the model have been explained, as well as the different values to be compared, a series of comparative tables between the values obtained by SNA Standard Tools (UCINET, Cytoscape, Pajek, Gephi) and the system created in this research through SPARQL queries will be presented below.

The results of the Social Network Analysis using SNA Standard Tools and our web-based application that uses the ontology is represented in the following table (Table 5):

**Table 5**

Results of the Social Network Analysis carried out using SNA Standard Tools and using KBS

| SNA Measures tested | Equivalent results |
|---|---|
| Friendship network measurements (density, members, isolated…) | Yes |
| Classmate network measurements | Yes |
| Acquaintance network measurements | Yes |
| Measurements of members in a friendship network | Yes |
| Measurements of members in a classmate network | Yes |
| Measurements of members in an acquaintance network | Yes |

## 6. Discussion:

Once the different comparative tables of results from the particular case study have been obtained, the system generated as a result of the ontology developed throughout this research, shows coherent and correct results when carrying out a social network analysis on three different networks.

These data are those used by the application generated and explained previously in this document, thanks to which, not only are the measures obtained at the SNA level, but real conclusions and an interpretation of the measures obtained in the SNA are obtained.

Thanks to the conclusions resulting from the implementation of the application used, the expert in drawing up the questions can, without having knowledge of SNA, obtain an interpretation of one or several social networks without needing knowledge of Social Network Analysis.

For this reason, this tool seems to facilitate the collection of data and their analysis in the field of research and in different domains, but especially in the field of medicine. Although it still needs improvements and solutions to small operating errors, it really fulfills the purpose of improving the handling of data by researchers and saves time both for the researchers and the respondents since it offers the opportunity to carry out surveys from any place where there is an Internet access point.

## 7. Conclusions:

As final conclusions to this research, we can say that it has been possible to define a conceptual model within the scope of Social Network Analysis, which in turn is adaptable to any domain of structured content in which an SNA is applied. This has been possible using semantic technologies, learning techniques and information filtering processes within the Semantic Web.

Furthermore, a development environment has been designed and developed that allows semantic technologies to be applied and making use of the conceptual model generated as the basis of a Knowledge Based System useful for multiple domains.

In a new step in the research, the generated conceptual model has been implemented, making use of the knowledge base generated on the basis of a series of data obtained through surveys or questionnaires and able to reach conclusions thanks to a series of SPARQL queries on the said model.

Based on some data, it has been possible to draw conclusions on an SNA in the field of health. This has been possible thanks to the rules and queries generated for that particular domain. However, the system allows the adaptation or use of the said system in any other domain, by previously inserting the specification of the particular knowledge for the domain in which it is required to apply the said System Based on Knowledge.

In this research, it has also been possible to obtain a characterization of the structured and semantic content common to any SNA, and which can also be applied indifferently in any domain.

Finally, thanks to the generation of semantic rules in the SWRL language, applying these with a semantic reasoning, in the case of this particular Pellet research, as well as thanks to the SPARQL queries, it has been possible to obtain the necessary base of an SBC to generate conclusions from a different Social Network Analysis, without the need to apply it to a single domain.

**Disclosure**

The author(s) declare(s) that there is no conflict of interest regarding the publication of this paper.

**Acknowledgment**

The work presented in this paper was supported by Junta de Castilla y León [grant number LE014G18].

**Competing Interests**

The authors declare that they have no conflicts of interests.

**References**

[1] S.P. Borgatti, M.G. Everett, Network analysis of 2-mode data, Soc. Networks. 19 (1997) 243–269. https://doi.org/10.1016/S0378-8733(96)00301-2.

[2] J. Leonidas Aguirre, Introducción al Análisis de Redes Sociales, Doc. Trab. CIEPP. (2011) 59.

[3]     E. Fernández-Martínez, E. Andina-Díaz, R. Fernández-Peña, R. García-López, I. Fulgueiras-Carril, C. Liébana-Presa, Social Networks, Engagement and Resilience in University Students, Int. J. Environ. Res. Public Health. 14 (2017) 1488. https://doi.org/10.3390/ijerph14121488.

[4]     J.A. Schneider, T. Walsh, B. Cornwell, D. Ostrow, S. Michaels, E.O. Laumann, HIV health center affiliation networks of black men who have sex with men: disentangling fragmented patterns of HIV prevention service utilization., Sex. Transm. Dis. 39 (2012) 598–604. https://doi.org/10.1097/OLQ.0b013e3182515cee.

[5]     A.M. Niekamp, L.A.G. Mercken, C.J.P.A. Hoebe, N.H.T.M. Dukers-Muijrers, A sexual affiliation network of swingers, heterosexuals practicing risk behaviours that potentiate the spread of sexually transmitted infections: A two-mode approach, Soc. Networks. 35 (2013) 223–236. https://doi.org/10.1016/j.socnet.2013.02.006.

[6]     E. Quiroga, I. García, J. Benítez-Andrades, C. Benavides, V. Martín, P. Marqués-Sánchez, A Qualitative Study of Secondary School Teachers' Perception of Social Network Analysis Metrics in the Context of Alcohol Consumption among Adolescents, Int. J. Environ. Res. Public Health. 14 (2017) 1531. https://doi.org/10.3390/ijerph14121531.

[7]     R.– D. Leon, R. Rodríguez-Rodríguez, P. Gómez-Gasquet, J. Mula, Social network analysis: A tool for evaluating and predicting future knowledge flows from an insurance organization, Technol. Forecast. Soc. Change. 114 (2017) 103–118. https://doi.org/10.1016/j.techfore.2016.07.032.

[8]     F.F. Ishtaiwa, I.M. Aburezeq, The impact of Google Docs on student collaboration: A UAE case study, Learn. Cult. Soc. Interact. 7 (2015) 85–96. https://doi.org/10.1016/j.lcsi.2015.07.004.

[9]     M. Huisman, M. a. J. Van Duijn, Software for Social Network Analysis, Model. Methods Soc. Netw. Anal. (2005) 270–316. https://doi.org/10.1017/CBO9780511811395.013.

[10]    L.F. S. Borgatti, M. Everett, Ucinet for windows: software for social network analysis, Anal. Technol. Harvard, MA. (2002).

[11]    G. Erétéo, M. Buffa, F. Gandon, P. Grohan, P. Sander, A state of the art on Social Network Analysis and its applications on a Semantic Web, Proc. SDoW2008 (Social Data Web), Work. Held with 7th Int. Semant. Web Conf. (2008) 1–6. http://citeseerx.ist.psu.edu/viewdoc/download?doi=10.1.1.142.7288&rep=rep1&type=pdf.

[12]    J.M. Ruiz-Martínez, R. Valencia-García, R. Martínez-Béjar, A. Hoffmann, BioOntoVerb: A top level ontology based framework to populate biomedical ontologies from texts, Knowledge-Based Syst. 36 (2012) 68–80. https://doi.org/10.1016/J.KNOSYS.2012.06.002.

[13]    S. Awaworyi Churchill, L. Farrell, Alcohol and depression: Evidence from the 2014 health survey for England, Drug Alcohol Depend. 180 (2017) 86–92. https://doi.org/10.1016/J.DRUGALCDEP.2017.08.006.

[14]    B. Baffour, T. Roselli, M. Haynes, J.J. Bon, M. Western, S. Clemens, Including mobile-only telephone users in a statewide preventive health survey—Differences in the prevalence of health risk factors and impact on trends, Prev. Med. Reports. 7 (2017) 91–98. https://doi.org/10.1016/J.PMEDR.2017.05.009.

[15]    K. Kwan, V. Do-Reynoso, G. Zarate-Gonzalez, S. Goldman-Mellor, Development and implementation of a community health survey for public health accreditation: Case study

from a rural county in California, Eval. Program Plann. 67 (2018) 47–52. https://doi.org/10.1016/J.EVALPROGPLAN.2017.11.004.

[16] A. Gordon, SurveyMonkey.com—Web-Based Survey and Evaluation System: http://www.SurveyMonkey.com, Internet High. Educ. 5 (2002) 83–87. https://doi.org/10.1016/S1096-7516(02)00061-1.

[17] M.-T. Cortés-Tomás, J.-A. Giménez-Costa, P. Motos-Sellés, M.-D. Sancerni-Beitia, Different versions of the Alcohol Use Disorders Identification Test (AUDIT) as screening instruments for underage binge drinking, Drug Alcohol Depend. 158 (2016) 52–59. https://doi.org/10.1016/j.drugalcdep.2015.10.033.

(2014) 209–215. https://doi.org/10.1016/j.drugalcdep.2014.06.017.

[18] M.J. Pardo-Guijarro, B. Woll, P. Moya-Martínez, M. Martínez-Andrés, E.E. Cortés-Ramírez, V. Martínez-Vizcaíno, Validity and reliability of the Spanish sign language version of the KIDSCREEN-27 health-related quality of life questionnaire for use in deaf children and adolescents, Gac. Sanit. 27 (2013) 318–324. https://doi.org/10.1016/j.gaceta.2012.11.003.

[19] Morris JF, Deckro RF (2013) SNA data difficulties with dark networks. Behav Sci Terror Polit Aggress 5:70–93. https://doi.org/10.1080/19434472.2012.731696 .

[20] A. Sapountzi, K.E. Psannis, Social networking data analysis tools challenges, Futur. Gener. Comput. Syst. (2016). doi:10.1016/j.future.2016.10.019 .

[21] T.J. Carney, C.M. Shea, Informatics Metrics and Measures for a Smart Public Health Systems Approach : Information Science Perspective, 2017 (2017).

[22] S. Kirrane, S. Villata, M. d'Aquin, Privacy, security and policies: A review of problems and solutions with semantic web technologies, Semant. Web. 9 (2018) 153–161. https://doi.org/10.3233/SW-180289.

[23] M. Elfida, M.K.M. Nasution, O.S. Sitompul, Enhancing to method for extracting Social network by the relation existence, IOP Conf. Ser. Mater. Sci. Eng. 300 (2018) 012057. https://doi.org/10.1088/1757-899X/300/1/012057.

[24] M.K.M. Nasution, O.S. Sitompul, S.A. Noah, Social network extraction based on Web: 3. the integrated superficial method, J. Phys. Conf. Ser. 978 (2018) 012033. https://doi.org/10.1088/1742-6596/978/1/012033.

[25] J. Kim, M. Hastak, Social network analysis: Characteristics of online social networks after a disaster, Int. J. Inf. Manage. 38 (2018) 86–96. https://doi.org/10.1016/J.IJINFOMGT.2017.08.003.

[26] S. Stieglitz, M. Mirbabaie, B. Ross, C. Neuberger, Social media analytics – Challenges in topic discovery, data collection, and data preparation, Int. J. Inf. Manage. 39 (2018) 156–168. https://doi.org/10.1016/J.IJINFOMGT.2017.12.002.

[27] B. Nie, S. Sun, Knowledge graph embedding via reasoning over entities, relations, and text, Futur. Gener. Comput. Syst. 91 (2019) 426–433. https://doi.org/10.1016/J.FUTURE.2018.09.040.

[28] A. Khaled, S. Ouchani, C. Chohra, Recommendations-based on semantic analysis of social networks in learning environments, Comput. Human Behav. (2018). https://doi.org/10.1016/J.CHB.2018.08.051.

[29] P.S. Raji, S. Surendran, RDF approach on social network analysis, in: 2016 Int. Conf. Res. Adv. Integr. Navig. Syst., IEEE, 2016: pp. 1–4. https://doi.org/10.1109/RAINS.2016.7764416.

[30] A. El Kassiri, F.-Z. Belouadha, Towards a unified semantic model for online social networks analysis and interoperability, in: 2015 10th Int. Conf. Intell. Syst. Theor. Appl., IEEE, 2015: pp. 1–6. https://doi.org/10.1109/SITA.2015.7358438.

[31] G. Ereteo, F. Gandon, M. Buffa, SemTagP: Semantic Community Detection in Folksonomies, in: 2011 IEEE/WIC/ACM Int. Conf. Web Intell. Intell. Agent Technol., IEEE, 2011: pp. 324–331. https://doi.org/10.1109/WI-IAT.2011.98.

[32] G. Ereteo, M. Buffa, F. Gandon, P. Grohan, M. Leitzelman, P. Sander, A State of the Art on Social Network Analysis and its Applications on a Semantic Web, (2008). https://hal.archives-ouvertes.fr/hal-00562064/ (accessed June 8, 2019).

[33] G. Erétéo, M. Buffa, F. Gandon, O. Corby, Analysis of a Real Online Social Network Using Semantic Web Frameworks, in: Springer, Berlin, Heidelberg, 2009: pp. 180–195. https://doi.org/10.1007/978-3-642-04930-9_12.

[34] G. Ereteo, Semantic Social Network Analysis, Telecom ParisTech, 2011. https://tel.archives-ouvertes.fr/tel-00586677.

[35] C. Welty, N. Guarino, Supporting ontological analysis of taxonomic relationships, Data Knowl. Eng. 39 (2001) 51–74. https://doi.org/10.1016/S0169-023X(01)00030-1.

[36] A. Gomez-Perez, V.R. Benjamins, Applications of Ontologies and Problem-Solving Methods, AI Mag. 20 (1999) 119. https://doi.org/10.1609/AIMAG.V20I1.1445.

[37] O. Cairó, S. Guardati, The KAMET II methodology: Knowledge acquisition, knowledge modeling and knowledge generation, Expert Syst. Appl. 39 (2012) 8108–8114. https://doi.org/10.1016/J.ESWA.2012.01.155.

[38] F.B. Ruy, G. Guizzardi, R.A. Falbo, C.C. Reginato, V.A. Santos, From reference ontologies to ontology patterns and back, Data Knowl. Eng. 109 (2017) 41–69. https://doi.org/10.1016/J.DATAK.2017.03.004.

[39] F.J. García-Peñalvo, R. Colomo-Palacios, J. García, R. Therón, Towards an ontology modeling tool. A validation in software engineering scenarios, Expert Syst. Appl. 39 (2012) 11468–11478. https://doi.org/10.1016/J.ESWA.2012.04.009.

[40] C. Gonzalez-Perez, B. Henderson-Sellers, T. McBride, G.C. Low, X. Larrucea, An Ontology for ISO software engineering standards: 2) Proof of concept and application, Comput. Stand. Interfaces. 48 (2016) 112–123. https://doi.org/10.1016/J.CSI.2016.04.007.

[41] W.W. Sim, P. Brouse, Towards an Ontology-based Persona-driven Requirements and Knowledge Engineering, Procedia Comput. Sci. 36 (2014) 314–321. https://doi.org/10.1016/J.PROCS.2014.09.099.

[42] L. Liu, P. Zhang, R. Fan, R. Zhang, H. Yang, Modeling ontology evolution with SetPi, Inf. Sci. (Ny). 255 (2014) 155–169. https://doi.org/10.1016/j.ins.2013.07.017.

[43] M. Reda Bouadjenek, H. Hacid, M. Bouzeghoub, Social Networks and Information Retrieval, How Are They Converging? A Survey, a Taxonomy and an Analysis of Social Information Retrieval Approaches and Platforms, Inf. Syst. 56 (2015) 1–18. https://doi.org/10.1016/j.is.2015.07.008.

[44] M. Alipour-Aghdam, Ontology-Driven Generic Questionnaire Design, (2014) 85.

[45] L. Lezcano, M.-A. Sicilia, C. Rodríguez-Solano, Integrating reasoning and clinical archetypes using OWL ontologies and SWRL rules, J. Biomed. Inform. 44 (2011) 343–353. https://doi.org/10.1016/J.JBI.2010.11.005.

[46] O. V. Popik, T. V. Ivanisenko, O. V. Saik, E.D. Petrovskiy, I.N. Lavrik, V.A.

Ivanisenko, NACE: A web-based tool for prediction of intercompartmental efficiency of human molecular genetic networks, Virus Res. 218 (2016) 79–85. https://doi.org/10.1016/j.virusres.2015.11.029.

[47] A.P. (FATEC) Paulino, G. (UFG) Fleuri, R. (USP) Zerbini, Application of the Louvain Community Detection, (2014).

[48] J.A. Benítez, J.E. Labra, E. Quiroga, V. Martín, I. García, P. Marqués-Sánchez, C. Benavides, A Web-Based Tool for Automatic Data Collection, Curation, and Visualization of Complex Healthcare Survey Studies including Social Network Analysis, Comput. Math. Methods Med. 2017 (2017) 1–8. https://doi.org/10.1155/2017/2579848.

[49] J. Powell, M. Hopkins, J. Powell, M. Hopkins, 18 – Drawing and serializing graphs, in: A Libr. Guid. to Graphs, Data Semant. Web, 2015: pp. 153–166. https://doi.org/10.1016/B978-1-84334-753-8.00018-X.

[50] M. Dabkowski, R. Breiger, F. Szidarovszky, Simultaneous-direct blockmodeling for multiple relations in Pajek, Soc. Networks. 40 (2015) 1–16. https://doi.org/10.1016/j.socnet.2014.06.003.